\documentclass[aps,prl,twocolumn,superscriptaddress,showpacs]{revtex4}
\usepackage{graphicx}

\begin{document}

\title{Phonon Effects on Spin-Charge Separation in One Dimension}
\author{Wen-Qiang Ning}
\affiliation{Department of Physics, Fudan University, Shanghai
200433, China} \affiliation{Department of Physics and Institute of
Theoretical Physics, The Chinese University of Hong Kong, Shatin,
Hong Kong, China}
\author{Hui Zhao}
\affiliation{Department of Physics, Fudan University, Shanghai
200433, China}
\author{Chang-Qin Wu }
 \email[Corresponding author. Email: ] {cqw@fudan.edu.cn}
\affiliation{Department of Physics, Fudan University, Shanghai
200433, China}
\author{Hai-Qing Lin}
\affiliation{Department of Physics and Institute of Theoretical Physics,
The Chinese University of Hong Kong, Shatin, Hong Kong, China}
\affiliation{Department of Physics, Fudan University, Shanghai 200433,
China}

\date{\today}

\begin{abstract}
Phonon effects on spin-charge separation in one dimension are investigated
through the calculation of one-electron spectral functions in terms of the
recently developed cluster perturbation theory together with  an optimized
phonon approach. It is found that the retardation effect due to the
finiteness of phonon frequency suppresses the spin-charge separation and
eventually makes it invisible in the spectral function. By comparing our
results with experimental data of TTF-TCNQ, it is observed that the
electron-phonon interaction must be taken into account when interpreting
the ARPES data.
\end{abstract}

\maketitle

It is commonly accepted that most one-dimensional (1D) correlated
electronic systems cannot be properly described by the traditional
Fermi liquid theory, instead, their behaviors are well predicted
by the Luttinger liquid theory
\cite{luttinger,lieb,emery,haldane}. One of the key features of
the Luttinger liquid is the spin-charge separation: the low-energy
excitations are not quasiparticles with charge $e$ and spin 1/2
together, rather, they are collective modes of spin and charge
excitations separately, called spinons and holons. Since spinon
and holon move with different speed, they eventually decouple.
Following earlier works \cite{luttinger,lieb,emery,haldane}, many
studied spin-charge separation with various theoretical schemes.
In particular, one could explore the existence of the spin-charge
separation by calculating the spectral function \cite{meden and
voit,meden,schrieffer,cpt1,Jeckelmann,mat}, which has direct
relation to the angle-resolved photoemission spectroscopy (ARPES).
In some recent works, the spinon and holon branches have been
observed by ARPES performed on some 1D materials such as
$SrCuO_2$.\cite{kim,kobayashi,claessen,sing} Because both
electron-electron and electron-phonon interactions exist in many
low-dimensional materials, it is important to address the role of
these interactions on the spin-charge separation.
\cite{meden,tsutsui,kivelson,perfetti,dessau,kmshen}

The one-dimensional Holstein-Hubbard model (HHM), which is the simplest
model involving both electron-phonon (e-p) and electron-electron (e-e)
interactions, has been used extensively to describe some low-dimensional
materials. Since electrons in these materials are strongly correlated, the
interplay between electron-phonon interaction and Coulomb repulsion should
have profound effect on the spin-charge separation, and we expect to
observe these effects by investigating the single-particle excitation
spectra. The spectral function provides valuable insights into the usually
complicated many-body systems, such as high-temperature superconductors,
cuprate ladder compounds, and organic conductors. For example, very
recently, by using the exact diagonalization method, Fehske \emph{et
al.}\cite{fehske} calculated the spectral function of the Holstein-Hubbard
model on a finite system and found a Mott-insulator to Peierls-insulator
transition at a compatible ratio of the e-e to e-p interactions.

Bearing these in mind, we compute the one-electron spectral function of the
HHM by applying the recently developed cluster perturbation theory
(CPT)\cite{cpt1,cpt2,hohenadler} together with an optimized phonon
approach.\cite{zhang,weise,zhao} The spectral function \emph{at full
frequency region} with rich satellite structures is obtained in the model
of \emph{both} e-e and e-p interactions for the first time. Phonon effects
on spin-charge separation are focused in the presence of e-e interactions
from weak to strong coupling. It is found that the retardation effect due
to the finiteness of phonon frequency does not favor the spin-charge
separation. In weak interaction regimes, a peak in the spectral function
was observed which is consistent with the existence of a metallic phase as
proposed recently by Clay and Hardikar\cite{clay}. Furthermore, it is
observed that one must take electron-phonon interaction into account when
interpreting the ARPES experimental data in the one dimensional material.

The HHM accounts for a tight-binding electron band, on-site
Coulomb repulsion between electrons of opposite spin, and coupling
of charge carriers to local phonons:
    \begin{eqnarray}
        H&=&-t\sum_{i,\sigma}(c_{i,\sigma}^\dag c_{i+1,\sigma}+H.c.)
        +U\sum_i n_{i\uparrow}n_{i\downarrow}\nonumber\\
        &&-g\sum_{i,\sigma}(b_i^\dag+b_i)n_{i,\sigma}
        +\omega_0\sum_ib_i^\dag b_i,
    \label{Model}
    \end{eqnarray}
where $c_{i,\sigma}^\dagger (c_{i,\sigma})$ creates (annihilates) an
electron with $\sigma$ on site $i$, and $b_i^\dag$ and $b_i$ are creation
and annihilation operators of the local phonon mode at site $i$,
respectively. $t$ is the electron hopping constant between nearest neighbor
sites which will be set as the energy unit in our calculations, $\omega_0$
is the bare phonon frequency, and $g$ is the electron-phonon coupling
constant.

For the calculation of the spectral properties within the framework of
CPT\cite{cpt1,cpt2}, one divides the lattice into fully equivalent clusters
of a finite sites. For each cluster, we calculate the Green's function
$G_{i,j}(z)(\equiv G_{i,j}^+(z)+G_{i,j}^-(z))$, with $G_{i,j}^{\pm}(z)$
defined as
\begin{equation}
G_{i,j}^{\pm}(z)=\langle\phi_0|c^{\pm}_i\frac{1}{z\pm
(H-E_0)}c_j^{\mp}|\phi_0\rangle,
\end{equation}
where $c^+_i\equiv c^\dagger_i, c^-_i\equiv c_i$, and $|\phi_0\rangle$
being the ground state of the cluster, which is obtained by using the
Lanczos exact diagonalization (ED) method within an optimized phonon
approach\cite{zhang,weise} under open boundary conditions. Two terms in
$G_{i,j}$ corresponding to electron and hole propagation, respectively, can
be obtained. The CPT treats the intercluster hopping by a strong-coupling
perturbation, i.e., ($t/U$) expansion. The lowest-order CPT approximation
to Green's function gives
\begin{equation}
G_{CPT}(k,z)=\frac{1}{N}\sum_{i,j}e^{-ik(i-j)}\widetilde{G}_{i,j}(Nk,z),
\end{equation}
where $\widetilde{G}_{i,j}(Q,z)$ is the Green's function of the full system
and $N$ is the size of clusters. The spectral function is then
$A(k,\omega)=-\frac{1}{\pi}Im[G_{CPT}(k,z)]$, where $z=\omega+i\eta$ with
$\eta$ defines the width of peaks in the spectral function. Since the Fermi
energy is set to zero, the spectral function has the symmetry of
$A(k,\omega)=A(\pi-k,-\omega)$ due to the electron-hole symmetry of the
model (1).

In the absence of interactions, $A(k,\omega)$ obtained by the CPT
method is exact \cite{cpt1,cpt2}, while for interacting models,
the accuracy of CPT depends on the size of the cluster and the
number of optimal phonon chosen. To test the accuracy of the
approach we use, we calculated the spectral function of the
Hubbard model with exactly the same parameters as used by Benthien
et al\cite{Jeckelmann} and obtained agreeable results. We also
calculated the first two spectral momenta and they match exactly
to those obtained by White\cite{white}. Furthermore, based on our
previous technique analysis\cite{zhao}, system parameters were
carefully chosen in this work to ensure that our results mimic
thermodynamic limit. Results obtained in this Letter were for
$N=6$, $\eta=0.1t$, and three optimized phonons at each site, with
relative error $10^{-5}$ for the total energy.\cite{zhao}

Three energy scales govern the physics of the HHM: the on-site Coulomb
repulsion ($U$), the electron-phonon coupling ($\lambda=2g^2/\omega_0$) and
the bare phonon frequency ($\omega_0$). The ground state of the system at
half-filling is a Mott-Hubbard insulating (MI) state when $U$ is large, and
shows spin-density-wave (SDW) fluctuations. When electron-phonon
interaction dominates, the system is in the Peierls insulating state (PI),
characterized by the charge-density-wave where both spin gap and change gap
are finite, while in the MI state, the spin gap vanishes. Very recently, it
was reported that there is a metallic region intermediate between the PI
and MI states with superconducting pairing correlation
dominates.\cite{clay}

In the absence of phonons ($g=0$), Eq. (\ref{Model}) is just the Hubbard
model whose physics have been extensively studied and it is well known that
in the Hubbard model, spin and charge separate.
\cite{schrieffer,cpt1,Jeckelmann,mat} On the other hand, when $U=0$, Eq.
(\ref{Model}) is another extensively studied model, the Holstein model
(HM). In the strong electron-phonon coupling region, the ground state of
the Holstein model at half filling is either a bipolaron insulating (BPI)
state in the large $\omega_0$ limit, characterized by the configuration
where each site is either empty or doubly occupied because phonons produce
an attraction between the electrons, or a traditional band insulating state
in the small $\omega_0$ limit.\cite{fehske,hirsch} In the weak coupling
region, the Peierls gap is suppressed by the phonon quantum fluctuations
and the ground state is at metallic (M) phase.\cite{wu,zhang,zhao} When the
phonon degrees of freedom are integrated out, the spin-1/2 Holstein model
could be mapped onto the Hubbard model with an effective dynamical
attraction $U_{eff}(\omega)=-\lambda/(1-\omega^2/\omega_0^2)$. Here one
also expects to observe the spin-charge separation. In the anti-adiabatic
limit, i.e., $\omega_0\rightarrow\infty$, the attraction becomes
instantaneous and equals to the bipolaron binding energy $\lambda$, which
has already been reported long time ago.\cite{hirsch} Obviously, for any
finite phonon frequency $\omega_0$, one must consider the retardation
effect fully which could not be simply presented by the above
$U_{eff}(\omega)$.

\begin{figure}
\includegraphics[scale=0.35]{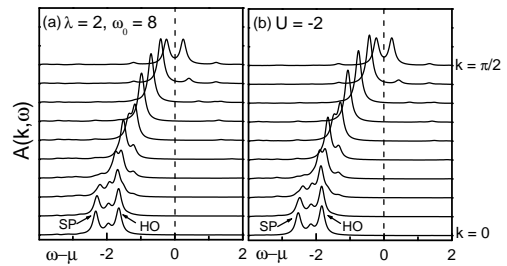}
\caption{The obtained spectral function $A(k,\omega)$ of (a) the
large-$\omega_0$ Holstein model and (b) the negative-U Hubbard
model at half-filling. HO and SP stand for holon and spinon
excitations respectively.\label{fig1}}
\end{figure}

To have a sense of the magnitude of $\omega_0$, we give a comparison of the
spectral function of these two models in Figure~1. It is clearly shown that
the spectral function of the Holstein model is almost the same as that of
the negative-U Hubbard model. Some minor differences due to finiteness of
$\omega_0$ are invisible in the figure (in other words, $\omega_0 = 8$ is
almost at the antiadibatic limit). This result is not trivial as seen at
first glance because it implies the single-particle excitation of the
system with a large phonon frequency is similar to that at the
antiadiabatic limit, which is consistent with the existence of a quantum
metal-insulator phase transition in the Holstein model.\cite{wu,jec,clay}
The peaks labelled ``SP'' and ``HO'' in Fig.~1 refer to the spinon and the
holon branches, respectively, signaturing the spin-charge separation.
Compare to the conventional Luttinger liquids (e.g., the positive-U Hubbard
model), the charge velocity ($v_\rho$) is smaller than the spin velocity
($v_\sigma$).

\begin{figure}
\includegraphics[scale=0.38]{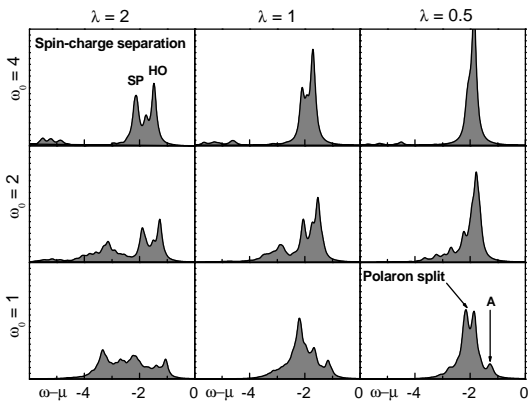}
\caption{The spectral function $A(0,\omega)$ of the Holstein
model. \label{fig2}}
\end{figure}

Fig.~2 illustrates the retardation effects systematically. Starting from
the strong coupling case ($\lambda = 2$), we observe that as we decrease
$\omega_0$ from 4 to 2, the incoherent part of the spectra becomes more
important. As a consequence, the spectral weights of the spinon and holon
excitations are much smaller. This could be regarded as a continuation from
Fig. 1: $\omega_0=\infty\rightarrow 8 \rightarrow 4$. When the phonon
frequency is further reduced, the spectral weights correspond to phonon
excitations become dominant, which is quite different from that in the
antiadiabatic regime where the ``spinon'' and ``holon'' excitations are
clearly the dominant ones. Therefore, due to the strong mixing of the
coherent and incoherent excitations, it is difficult to single out the
spinon and holon excitations, instead, one observes an almost flat band
dispersion with exponentially small spectral weight. The dominant peaks in
the incoherent part of the spectra are related to multiples of the (large)
bare phonon frequency broadened by electronic excitations. Such
electron-phonon mixed nature of excitations could be seen in the spectra
away from the Fermi surface.

It is quite natural to expect that the separation of spin and charge
excitation will become smaller with the decrease of the electron-phonon
coupling strength. This is clearly reflected in the spectral function. As
shown in the first row of Fig. 2, when we reduce the electron-phonon
coupling $\lambda$ at $\omega_0=4$, the difference between the spinon and
the holon excitation at $k=0$ becomes smaller and eventually invisible. As
we reduce phonon frequency, retardation comes into play. It is also
interesting to observe that there is an excitation split in the weak
coupling case (Fig.~2, $\lambda=0.5$, $\omega_0=1$). Such splitting is not
due to spin-charge separation. In fact, by carefully comparing this
spectral function with that of a spinless Holstein model at corresponding
electron-phonon coupling, we found that the splitting is caused by the
polaron interaction.

\begin{figure}
\includegraphics[scale=0.35]{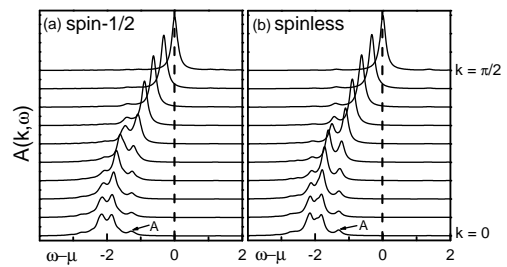}
\caption{Spectral function $A(k,\omega)$ of the spinful (a) and
spinless (b) Holstein model. $\lambda=0.5$ and $\omega_0=1$.
\label{fig3}}
\end{figure}

Figure 3 shows the spectra of up-spin electron for the spin-1/2 Holstein
model at half-filling in comparison with the spinless Holstein model. In
the weak coupling regime the two spectral are almost the same, indicating
that the existence of the down-spin electrons have nearly zero effect on
the spectral function of the up-spin electrons. It shows that the
phonon-mediated interaction between up- and down-spin electrons is very
weak so they are almost decoupled in this case. Thus the splitting of the
excitations cannot be attributed to the spin-charge separation, rather, it
is due to the interaction among polarons. Notice that there is a small peak
labelled as ``A'' in the spectral weight, which is almost dispersionless in
small $k$ regime (see Fig.~3) and is suppressed by the on-site repulsion
$U$ (see Fig.~4). Since it is appeared in the metallic region intermediate
between PI and MI phases, we speculate the peak ``A'' may be related with
the electron pairing, as discussed recently by Clay and
Hardiker\cite{clay}, although further investigations are definitely
deserved.

Now we turn the Coulomb interaction $U$ on and discuss its effect
on the spectral function in the presence of electron-phonon
interaction.
\begin{figure}
\includegraphics[scale=0.38]{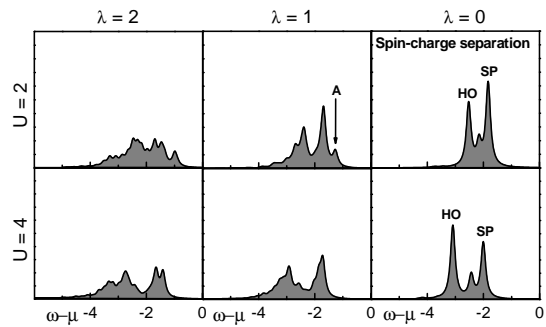}
\caption{Spectral function $A(0,\omega)$ of the Holstein-Hubbard
model at half filling. $\omega_0=1$.\label{fig4}}
\end{figure}
From Figure 4, we also see that, the electron-electron interaction
and the electron-phonon interaction have opposite effect on the
spin-charge separation, as we intuitively expected. At given
electron-electron interaction (for example, $U=4$), increasing the
electron-phonon interaction tends to broaden the excitation bands
and leave the spin-charge separation invisible. While at fix
electron-phonon interaction, the electron-electron interaction
increases the separation between the spin and the charge
excitations.

\begin{figure}
\includegraphics[scale=0.38]{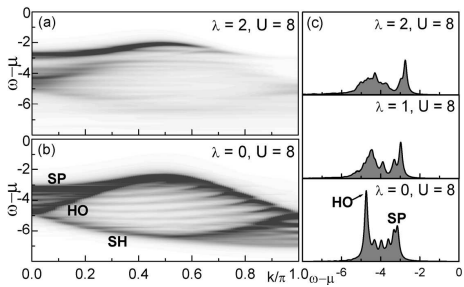}
\caption{Density plot of the spectral function $A(k,\omega)$ for
(a) the Holstein-Hubbard, and (b) the Hubbard model. (c) The
spectral function $A(0,\omega)$ of corresponding models at
different electron-phonon coupling. SH stands for the shadow band.
$\omega_0=1.$\label{fig5}}
\end{figure}

By further increasing the electron-electron interaction, we illustrate the
role of electron-phonon coupling in Fig.~5. Comparing Fig.~5(b) with
Fig.~5(a), one may still observe the signature of spin-charge separation as
$\lambda$ increases, but the holon branch is much more broadened and its
spectral weight decreases, while the spinon branch is nearly unchanged.
This is clearly shown in Fig.~5(c). Consequently, the so-called shadow
band, which originates from the diverging spin fluctuations at
$2k_F$,\cite{hass,penc,favand,cpt1} is nearly gone, due to the fact that
the shadow band is actually the continuation of the holon band.

Finally, we make a comparison on our results to the ARPES experiments of
the quasi-one-dimensional organic conductor TTF-TCNQ.\cite{claessen,sing}
We notice that in the Fig.~7 of the Ref.\cite{sing}, the charge branch (b)
and the spin branch (a) are weakly connected with each other. The result
can not be explained by the Hubbard model alone(Fig.~5(b)). As our results
suggest, the electron-phonon interaction must be taken into account. From
Fig. 5(a), it is clearly seen that the dispersion of the charge branch is
weakened in the spectral function and the spinon branch is weakly connected
to the holon branch. A similar broadening due to phonon is also observed in
one dimensional SrCuO$_2$.\cite{zxshen} These facts indicate that one
should expect significant contribution from the electron-phonon interaction
to the spectra of these strongly correlated quasi-one-dimensional
materials. Of course, our conclusions are based on numerical studies of Eq.
(1) so to have detailed analyzes of experiments, one may use models
different from Eq. (1) but we believe essential physics remain unchanged.

In summary, by applying the CPT together with an optimized phonon approach,
we have studied the spectral function of the one-dimensional
Holstein-Hubbard model at half filling. A comprehensive picture for the
spectral function in the presence of electron-electron interaction and
electron-phonon interaction was presented. In particular, we addressed the
issue of spin-charge separation and found that the electron-electron
interaction competes with the electron-phonon interaction on the
spin-charge separation, and the retardation effect due to phonons may
diminish the spin-charge separation in the spectral function. We also found
polaron splitting and observed a peak that may related to electron pairing
in the weak e-p coupling limit.

This work was partially supported by the National Natural Science
Foundation of China, CUHK 401504, and MOE B06011. The authors are
grateful to Prof. R. B. Tao for his generous support and to Prof.
D. L. Feng for helpful discussion on the ARPES data. We also thank
Profs. H. Chen and H. Zheng for stimulating discussions and Prof.
Z.-X. Shen for his preprint.

\end{document}